# Interactive Planning and Operations using Peak Load Pricing in Distribution Systems

**M. ILIC[1], M. GOUGH[1,2]**
Massachusetts Institute of Technology[1], University of Porto[2]
USA[1], Portugal[2]

**SUMMARY**

The emergence of Distributed Energy Resources (DERs) provides both challenges and opportunities for the planning and operations of distribution systems. These resources can be deployed in a manner that is either complementary to or in competition with traditional network operations and planning as the DERs can provide numerous important services to grid operators and utilities. The key to harnessing the full potential of these DERs in order to work with traditional network investments (such as line upgrades, and installation of network switches) is to accurately quantify the value that these different resources can offer to the system and thus estimate the trade-offs which occur when the investment decision is taken to use a certain technology. This paper presents a novel method to estimate this trade-off using a dynamic Peak-Load Pricing (PLP) methodology to quantify the costs and benefits of two distinct technologies (distributed generation and network switches). PLP is a pricing strategy for a time-dependent quantity of a non-storable commodity and is based on the theory of long-run marginal costs. Importantly PLP deals with the trade-off between capacity utilization and consumer welfare. Therefore, the capacity price is set at a point where the cost of investment is exactly offset by the additional social welfare that the investment would bring. Importantly it allows for capital cost recovery with no uplift payments. This dynamic PLP methodology is an interactive planning and operations model based on the Dynamic Monitoring and Decision Systems (DyMonDS) Framework which helps to align physical, information, and economic incentives across many stakeholders within the electric energy system. The DyMonDS framework helps to solve the drawbacks of PLP, which are related to the computational complexity of the PLP models considering different technologies with different payback periods over a long investment horizon. Additionally, this model helps to reduce the issue of the lumpiness of investment in network assets. This is achieved through the Model Predictive Control capabilities of DyMonDS to sequentially clear both the energy balance and the investment decisions per time step. Results show that the dynamic PLP model can accurately value the impact of different technologies in reducing congestion and increasing the number of customers served. This allows the network operator to easily identify which technologies should be chosen in each investment cycle. This model can be used by utilities, regulators, or aggregators of DERs to determine the accurate value of investments while guaranteeing capital cost recovery through the PLP-determined price.

mbgough@mit.edu

**NOMENCLATURE**

**Sets**
| | |
|---|---|
| g | Set of generators |
| l,i | Set of lines |
| n | Set of nodes |
| $t_{inv}$ | Set of time for investment |
| $t_{RT}$ | Set of time for real-time market |
| z | Set of loads |

**Parameters**
| | |
|---|---|
| $\kappa_t$ | Unit cost of capacity upgrade at time $t$ |
| $L_z^{max}$ | Maximum load for load $z$ |
| $P_{g,t}^{min}, P_{g,t}^{max}$ | Min and Max power output of asset $g$ at time $t$ |

**Variables**
| | |
|---|---|
| $\lambda(\hat{g})$ | Price of electricity from DER $g$ |
| $\lambda(\hat{s})$ | Fee paid to Aggregator/Utility for system upgrade |
| $\lambda_z(t)$ | Price of electricity for Load $z$ at time $t$ |
| $C_{g,t}$ | Cost of generation from DER $g$ at time $t$ |
| $K_t$ | Size of network upgrade for Aggregator/DSO determined by DSO |
| $P_{g,t}$ | Power generated by DER $g$ at time $t$ |
| $U_z(L_z(t))$ | Utility derived from load $z$ at time $t$ |

**INTRODUCTION**

Distributed energy resources (DERs) have the potential to play a key role in future energy systems. The proliferation of these devices, which can be owned and operated by individual consumers, has largely been driven by rapidly declining costs. While each DER can have a small impact on the wider grid, the ability to coordinate fleets of DERs means that these fleets can have significant potential to meet various objectives of system operators or consumers and play a significant role in future energy systems [1]. The increasing digitization of the energy system allows increasing levels of communication between, and control of, various devices to serve different needs. This increased communications capability is coupled with increasing embedded computational capacity within the DERs to further the ability to coordinate and control large numbers of DERs [2].

The impacts of DERs depend on the type, the number of assets controlled, and the location of the DER in relation to the wider distribution system. Additionally, these impacts can vary temporally. These fluctuations in potential impact (and thus fluctuations in value) mean that incorporating DERs into operational and planning models for DS is extremely challenging. The owners and operators may also have different objectives and preferences for installing the DERs and this variety imposes additional uncertainty and is difficult to capture using existing traditional centralized planning methodologies. DERs can provide numerous benefits to the power system including providing both energy and capacity generation value, ancillary services such as reserves, frequency regulation, and ramping support, and finally can also provide benefits related to the delivery of electricity such as acting as Non-Wires Alternatives (NWAs) and voltage support [3].

There is a growing host of research showing how DERs can be used as alternatives to traditional distribution system investments [4]. In these situations, DERs are referred to as NWAs. DERs can provide increased planning flexibility as they typically have shorter installation times, smaller investments, and can be modular. This helps reduce the risk of implementing distribution network upgrades to meet expected load growth in the future and then that load growth not materializing.

Agents who can coordinate and control large numbers of DERs to participate in energy markets as a single actor are poised to play a significant role in future distribution systems [5]. In a future distribution system that has many DERs actively participating, the means to aggregate a diverse set of DERs to operate as a single entity in energy markets will be essential. Having several aggregators bidding into the energy markets can increase competition as long as the aggregators are large enough to bid efficiently but not large enough that they can exercise market power [6]. The falling costs of DERs are reducing the barrier to entry for DER ownership but the barrier to entry into energy markets is still prohibitively high for individual DERs owners. Aggregators can help reduce this barrier to entry by grouping many small DERs, allowing them to bid into energy markets as a single entity.

In future distribution systems, competition for energy services may exist as consumers with different preferences will be able to choose the type and level of their energy service contracts. In such a world, ensuring a technology-neutral manner of valuing assets capable of providing these energy services would be crucial to the long-term efficiency and sustainability of energy services at the distribution level. There has been an increase in the research around the introduction of competition into distribution systems, not in terms of installing duplicate sets of lines and traditional infrastructure investments, but in terms of delivering energy services to consumers. This can be done by using a diverse set of DERs to provide energy services to consumers as well as system services to system operators [7]. This introduces interesting questions about open access, ownership rights, and how to ensure fair competition within distribution systems [6]. Within a newly competitive distribution system, DER aggregators will have a major role in managing and bundling together diverse assets to provide a robust portfolio capable of providing reliable energy services to maximize social welfare [8].

**CONTEXT**

Existing pricing strategies have relied on marginal pricing from classical economic theory, which, under certain assumptions, lead to efficient outcomes measured through the maximization of social welfare. This paper proposes that this approach fails to accurately value DERs, assets with high investment costs and low operation and maintenance costs, which can provide alternative benefits compared to traditional investments. In the case of DERs (typically renewable energy generators or battery energy storage systems), the marginal costs are low and are below the average cost. When applied in these circumstances, marginal cost pricing results in deficits that must be recovered. These are typically recovered through various taxes or uplift payments [9]. Furthermore, increasing DER deployment is expected to increase both the short-term and long-term price elasticity of demand [6]. This further supports the introduction of competition for energy services in the distribution system.

This paper provides a novel decision-making framework for distribution systems considering both short-term operations and long-term capacity upgrades using a dynamic peak load pricing approach. This initial model introduces competition into distribution systems for both energy delivery and infrastructure upgrades. The preferences of the consumers are internalized and included in the cost functions that they provide to the Distribution System Operator (DSO). These contributions allow the model to measure the trade-offs between different classes of network upgrades. This paper applies Peak Load Pricing to distribution system planning to accurately value DERs and DR. This ensures that the DERs and other potential distribution system upgrades are priced according to their long-run marginal costs (LRMC). The paper argues that due to the relatively low capital costs of DERs (which are expected to decrease further) combined with their modular nature, the problem of the lumpiness of investment is greatly minimized. In addition, the distributed decision-making framework used within the Dynamic Monitoring and Decision Systems (DyMonDS) framework reduces computational complexity and allows each agent within the system to make their own decisions regarding capacity investment or participation in possible energy markets.

Peak-load Pricing (PLP) was developed in the mid-twentieth century by Boiteux (1949) and Steiner (1957) [10]. The model was extensively developed by Michael Crew and Paul Kleindofer with strong contributions by Hung-po Chao [11]. PLP is a pricing strategy for a time-dependent quantity of a non-storable commodity and is based on the theory of long-run marginal costs (LRMC) [12]. As the demand varies with time, there is a need to invest in sufficient capacity to meet the peak demand, but this capacity is not typically used at non-peak times. Existing work has extended PLP to consider multiple technologies available to meet the demand as well as both supply and demand uncertainty [10]. PLP has been applied to electricity transmission planning by [12] and centralized distribution expansion planning [13]. Importantly in PLP, the budget balance constraint is considered directly and so PLP deals with the tradeoff between capacity utilization and consumer welfare. In addition, the distributed generation which is based on renewable energy technologies such as solar PV and wind, has most of the costs being capital costs with only a small fraction of the costs being related to operations or maintenance. PLP explicitly considers the capital costs, therefore, removing any distortions or requirements for uplift payments that may be required in other pricing strategies [9].

The DyMonDS framework has been developed over many years for various applications in the electric power system [2]. This framework has been proven as a computationally efficient and robust distributed decision-making framework for the electric power sector. Fundamentally, DyMonDS is a multi-layered cyber-physical representation of an electric power system. In the lower layers, agents are responsible for their decision-making regarding their objectives, bidding strategies, capacity expansion plans, and risk preferences. These decisions are internalized by the agents and the output of their decision-making process is a set of bid curves that are communicated to the upper layers, which in turn optimize the system in a distributed manner. A key aspect of DyMonDS is that the individual agents compute internal bid curves for production and elastic demand. This allows consumers or generators to internalize constraints and preferences into their cost functions. This strategy can help to differentiate groups of consumers or producers. The outcome of the DyMonDS process is a set of physically implementable bids and the system aligns both technical and economic signals [2].

**DYNAMIC PEAK LOAD PRICING**

This paper proposes a dynamic application of peak load pricing to distribution system operations and expansion planning. This is done to allow a diverse set of assets to bid into competitive markets to provide energy to consumers fairly and efficiently, based on their long-run marginal cost. The proposed model is built upon the work in [13] and [14]. As such, the model is developed as a hierarchical decision-making framework with the different agents located at different levels within the model and it is shown in Figure 1. The framework aims to quantify the trade-offs between traditional investments and DERs according to PLP to account for the LRMC of the assets.

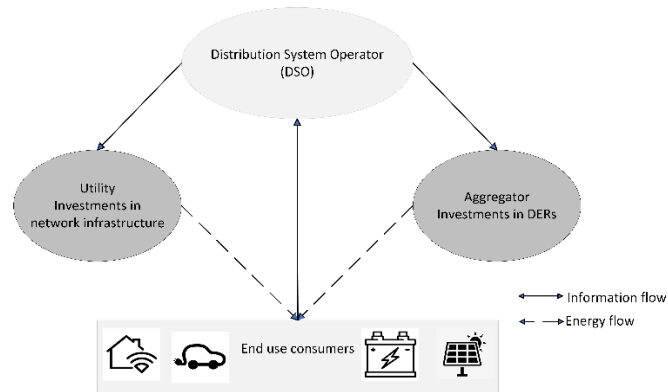

*Figure 1: Structure of model*

At the heart of the proposed model is the DSO. The DSO will act as a central clearing agent for both demand and supply bids submitted by consumers and DERs. The DSO will send out forecasts for both short- and long-term demand. The aggregators and utility submit both short-term and long-term bids to the DSO. The short-term bids are to meet the energy demand of the consumers while the long-term bids are for investment in network upgrades. The DSO has no financial incentives in this model but purely acts to settle the price and quantity of energy traded. The DSO also does not own any infrastructure. The DSO liaises with several aggregators and a single utility. These agents coordinate with the DSO in the upper level as well as consumers in the lower level. The agents in the middle layer are responsible for aggregating the demand of several small consumers from the lower level and then submitting bid curves to the DSO to meet the demand of the consumers. Aggregators may own DERs within the system and in this case, they also submit generation curves to the DSO to be eligible for dispatch. The aggregators may make investments in DERs depending on the long-term price signals provided by the DSO. The existing utility is also responsible for submitting bid curves to the DSO to meet the demand of its consumers. The utility may make investments in network infrastructure such as reinforcing lines or investing in switching devices. The utility may not invest in DERs in this proposed model.

Both aggregators and the utility will develop internal bid curves. In the case of aggregators with DERs, the offer curves will also be developed internally, which will include internal considerations of any constraints (generation, ramping) of the DERs. Following this framework, the only information that is exchanged between the DSO and the suppliers (both utility and aggregators) is the demand forecasts and the information sent from the suppliers to the DSO are sets of supply curves. Once the DSO has received both the short-term and long-term bid curves, it performs a dispatch using the simple convex bids provided by the DERs [2]. This reduces the amount of information that needs to be sent at each time step, thus increasing the computational efficiency of the model.

The DERs will receive the forecasts of the load and the price from the DSO. The DERs will then run their optimization to internally develop a supply curve according to their characteristics and then decide whether to bid into the market for this period. The suppliers may perturb their bid amount by a fixed amount to generate supply curves which are then sent to the DSO for clearing. Having the DERs individually generate these supply curves reduces the computational complexity of the model while also acting as a risk management measure. The DSO does not need to know detailed information from the suppliers such as ramp rates or minimum up/down times.

There are three components of the revenue for the Aggregators and Utilities. The first component is the revenues from selling electricity to consumers. The second component is the utility of satisfying the elastic load of the consumers presented in a financial sense. The final component is related to the revenues for undertaking infrastructure upgrades (in the case of the aggregators this is investing in DER capacity while in the case of the utility infrastructure upgrades are investments in switches) and these components are shown in Eq. (1). In this initial proposal, only the aggregator can invest in new DER capacity and only the Utility can invest in new switches. In future models, this division may shift so that either the aggregator or the utility can invest in different types of infrastructure upgrades.

$$\text{Max} \sum_{g,t_{RT},z}^{G,T_{RT},Z} [P_{g,t_{RT}}(\lambda(\hat{g}) - C_{g,t_{RT}}) + (U_z(L_z(t_{RT})) - \lambda(t)(L_z(t)))]$$
$$+ \sum_{t_{INV}}^{T_{INV}} [K_{t_{INV}}(\lambda(\hat{s}) - \kappa_{t_{INV}})] \quad (1)$$

In the first component of the revenue, the aggregator and utility attempt to maximize the amount of energy sold to customers minus the cost of producing that energy. The second component is related to the outcomes of the long-term bids for infrastructure upgrades that the utility and aggregators submit to the DSO. The various aggregators will compete directly with the utility to be awarded these network investments. The fees awarded by the DSO contain a per unit price of capacity expansion determined by the DSO but the investment in DER expansion is determined by the aggregator and this is shown in Eq. (1). Finally, a financial representation of the utility derived by the consumers for meeting their elastic load is shown in the third component of Eq. (1).

The utility competes with the aggregators to provide network upgrades. The utility along with the aggregators submits long-term investment bids to the DSO who allocates the investment capacity based on least cost planning while meeting the energy demand of the consumers. The investment bids submitted by the aggregator and utility are based on peak-load pricing to recover both capital and operating costs of network upgrades. The aggregator and utility also receive a unit revenue, shown as λ(s) for each unit of investment undertaken from the DSO. This investment has a unit cost of κ$_t$ in Eq. (1), for investing in new infrastructure upgrades. This determines the optimal location and capacity of these investments for a given period based on their long-term bids. A load of a consumer at a certain instance is bounded between 0 and the maximum load of that consumer.

$$0 \leq L_z \leq L_z^{max} \quad (2)$$

Subject to the following ramping and generation limits:

$$P_{t_{RT}}^{max}(G) = h_i\left(P_{t_{RT-1}}^{max}(G)\right) \tag{3}$$

$$P_{t_{RT}}^{min}(G) = g_i\left(P_{t_{RT-1}}^{min}(G)\right) \tag{4}$$

$$P_{g,t_{RT}}^{min} \leq P_{g,t_{RT}} \leq P_{g,t_{RT}}^{max} \tag{5}$$

The DSO aims to maximize social welfare over the long term. The DSO acts as a market clearing agent for the bids from the utility and the aggregators. The DSO also clears the bids for long-term system upgrades from the utility in terms of increased switches or the aggregators in terms of increased DER investment.

Considering this model, the power flows in the network under consideration are modeled using the DC power flow equations. These define the power flow in line $i,j$ as

$$T^t = \mathbf{H}P^t \quad \forall (t) \tag{6}$$

with **H** representing the $n_l * n$ transfer admittance matrix. The system's power balance is represented by:

$$\sum_g^G P_{g,t_{RT}} - \sum_l^L \sum_n^N P_{G,l,n} = 0 \quad \forall (t) \tag{7}$$

Following the work done by [12], this proposed model assumes a constant admittance matrix **H** and in doing so considers no direct relationship between line capacity and line reactance.

In this formulation, a network infrastructure upgrade of capacity K has a unit cost of κ. The upgrades are strictly positive, i.e.:

$$0 \leq K_{t_{INV}} \quad \forall (t) \tag{8}$$

Following the above equations, the problem is solved by forming the Lagrangian $\mathcal{L}$ as shown below:

$$\mathcal{L}(P_{g,t_{RT}}, K_{inv}) = \Omega(P_{g,t_{RT}}, K_{inv})$$
$$+ \sum_{l,t}^{n_l,T} \mu_{l,t}(K_l - \sum_i^N (H_{l,g} P_{g,t_{RT}})) + \sum_{l,t}^{n_l,T} \nu_{l,t}(H_{l,g} P_{g,t_{RT}} + K_l) + \sum_{t_{RT}}^{T_{RT}} \lambda_{t_{RT}} \sum_i^N P_{g,t_{RT}}$$
$$+ \sum_i^{n_l} \zeta_t K_t + \sum_{l,t_{RT}}^{n_l,T_{RT}} \alpha(P_{g,t_{RT}} - P_{min}) + \sum_{l,t_{RT}}^{n_l,T_{RT}} \beta(P_{max,t_{RT}} - P_{g,t_{RT}}) \tag{9}$$

This problem has been validated in [12] with optimality conditions resulting from the Karush–Kuhn–Tucker conditions. Assuming the power injection constraint shown in Eq. (5) and null Lagrange multipliers $\alpha_g$ and $\beta_g$, the following equations are obtained:

$$\sum_{g,t}^{G,T_{RT}} P_{g,t_{RT}} = 0 \tag{10}$$

$$\mu_{l,t} \geq 0 \text{ and } \mu_{l,t}(K_l - \sum_i^N (H_{l,g} P_{g,t_{RT}})) = 0 \quad \forall (l,t) \tag{11}$$

$$\nu_{l,t} \geq 0 \text{ and } \nu_{l,t} \sum_i^N (H_{l,g} P_{g,t_{RT}} + K_l) = 0 \quad \forall (l,t) \tag{12}$$

$$\lambda_{t_{RT}} = \rho_{g,t} + \sum_i^N H_{l,g}(\mu_{g,t} - \nu_{g,t}) \quad \forall (l,t) \tag{13}$$

$$\kappa_{t_{INV}} \geq \sum_t^T (\mu_{g,t} + \nu_{g,t}) \text{ and } K_{INV}\left[\sum_t^T (\mu_{g,t} + \nu_{g,t}) - \kappa_{t_{INV}}\right] = 0 \quad \forall (l,t) \tag{14}$$

The shadow prices of network expansion are represented by $\mu_{l,t}$ and $\nu_{l,t}$. Work done in [12] shows that while Eq. (14) has the same form of Short Run Marginal Cost-based power flow indices, the shadow prices, $\mu_{l,t}$ and $\nu_{l,t}$, represent capacity costs as opposed to congestion costs. This PLP formulation the allocates the network expansion in an optimal manner so that there is an exact balance between the costs of these expansions and the added social welfare that these expansions contribute [12].

The DSO uses this formulation to systematically clear the generation and load bids from the aggregators and the utility. The DSO follows a sequential process to clear both the long and short-term energy prices and communicates these prices to the Aggregators and the utility. By transmitting these prices, the DSO indicates which, if any, network upgrades are required to meet the reported demand. The process includes the following steps:
Step 1: The aggregators and utility submit their bid and demand curves to the DSO.
Step 2: The DSO uses information regarding the technical constraints of the network plus the bid functions from the aggregator and utility to find the generation and price that maximizes social welfare. The decision variables at this stage are the price and generation needed to meet the demand in the given time window. Within Step 2, the optimal solutions for capacity and price may not be found in the first iteration. If this is the case, new price points can be sent from the DSO to the aggregator and utility allowing them to update their bid curves.
Step 3: Once a solution to the iterative process in Step 2 is found, the DSO publishes the capacity and prices to the aggregators and utility. The process begins again for the next step in the time horizon.

The model introduced in the previous section was applied to a test system to quantify the extent to which DERs can compete with traditional utilities in providing energy services to consumers. The test system was the modified Roy Billington Test system as used in [14] and is shown in Fig. 3. Within the test systems, several candidate locations for the switches (both normally open (NOS) and normally closed (NCS)) and DERs were selected. The locations for the switches are shown in locations A-M and the locations for the DERs are locations 1 and 2 of Figure 2. At these points, the installed capacity of the DERs can vary. The consumers are aggregated at the various nodes. Consumers have different loads and value these loads differently by submitting differing demand curves. The costs of the network investments are taken from [14] and are shown in Table 1. A discount rate of 7% was used.

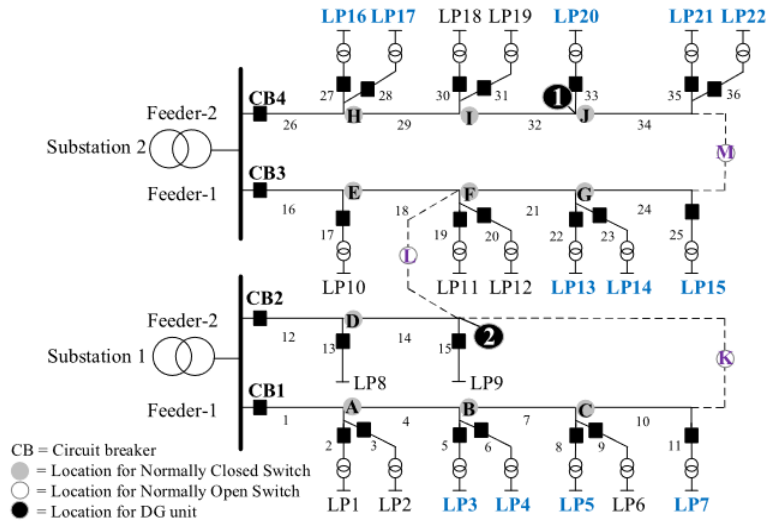

*Figure 2: Test system used*

*Table 1: Costs of network investments*

| Type of investment | Capital costs | Operating costs |
|---|---|---|
| NOS | $85 000 | $200/ year |
| NCS | $20 000 | $200/ year |
| DERs | $340/kW | $17/kW/ year |

**RESULTS**

The results of the case study show the final value of investments in either network switches or increases in DG capacity. The results of the investments in network switches are shown in Table 2. The table shows the clear benefit of installing new switches as the price decreases and the number of consumers served increases until the optimal number of switches is installed, which in this case is nine.

*Table 2: Results for network switches*

| Number of switches | Number of clients served | Price ($/MWh) | Location of switches |
|---|---|---|---|
| 4 | 1330 | 12.7235 | ABKM |
| 5 | 1530 | 11.7541 | ABCKM |
| 6 | 1710 | 11.0689 | ABCHKM |
| 7 | 1710 | 11.4156 | ABCDHKM |
| 8 | 1910 | 11.1073 | ABCFGKLM |
| 9 | 2090 | 10.6023 | ABCFGHKLM |
| 10 | 2090 | 10.8809 | ABCDFGHKLM |
| 11 | 2090 | 11.1596 | ABCDEFGHKLM |
| 12 | 2090 | 11.4383 | ABCDEFGHIKLM |
| 13 | 2090 | 11.717 | ABCDEFGHIJKLM |

The model highlights the most beneficial locations to install the switches. The additional number of customers served per additional switch and the impact of this on the final price is shown in Figure 3.

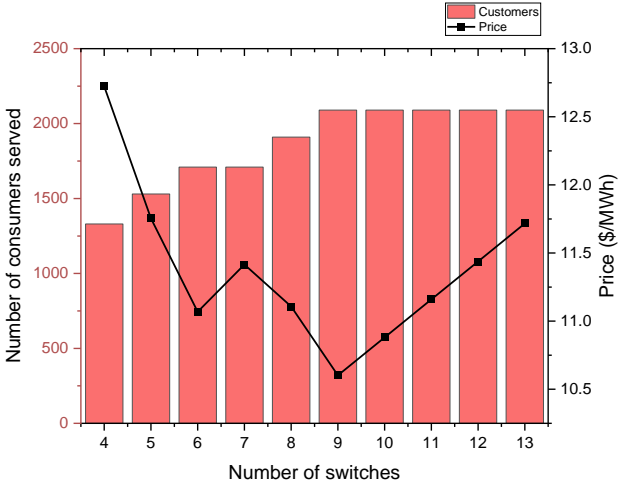

*Figure 3: Number of customers served*

Finally, the impact of increased DG investment on the price is shown in Figure 4. There is a decline in the price until a capacity of 1.2 MW is reached and then above this the price begins to increase as the excess capital costs of the capacity of DG needs to accounted for. The number of consumers served remains the same as the DG acts as supplemental generation for the existing network.

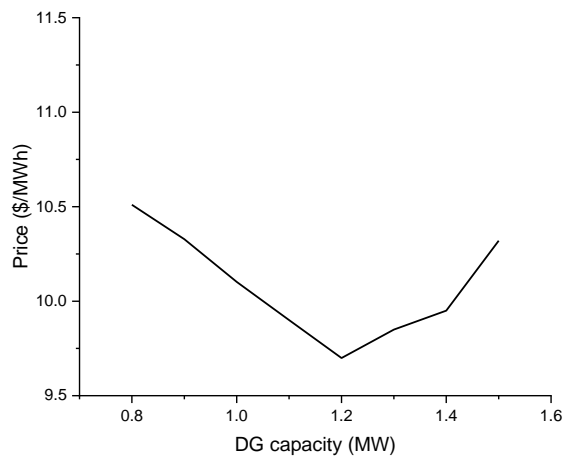

*Figure 4: Investments in DER capacity*

**CONCLUSIONS**

This paper presented an interactive operations and planning model for distribution systems using a modified version of the Peak Load Pricing method. The model was implemented within the DyMonDS framework in order to provide a robust and distributed decision-making process for the optimal investments in either network switches or distributed generation. Results show that the model provided the optimal investments for both technologies. Using this framework, the trade-offs between each technology can be easily and accurately calculated allowing for a transparent process of identifying investments in network expansion while operating the system in an optimal manner.

Currently, the model only investigates congestion management provided by two technologies. Future work can expand on the number of technologies considered and different services provided (voltage, frequency). Additionally, the impact of demand response should be included.

**ACKNOWLEDGEMENT**

The authors would like to thank Siripha Junlakarn of the Energy Research Unit at Chulalongkorn University, Thailand, for constructive discussions and for software provision. M. Gough was funded by a Ph.D. Scholarship from the *Fundação para a Ciência e a Tecnologia* with reference number UI/BD/152279/2021.

**BIBLIOGRAPHY**

[1] M. Obi, T. Slay, and R. Bass, "Distributed energy resource aggregation using customer-owned equipment: A review of literature and standards," *Energy Rep.*, vol. 6, pp. 2358–2369, Nov. 2020, doi: 10.1016/j.egyr.2020.08.035.
[2] M. D. Ilic, R. Jaddivada, and M. Korpas, "Interactive protocols for distributed energy resource management systems (DERMS)," *IET Gener. Transm. Distrib.*, vol. 14, no. 11, pp. 2065–2081, Jun. 2020, doi: 10.1049/iet-gtd.2019.1022.
[3] N. Mims Frick, S. Price, L. Schwartz, N. Hanus, and S. Ben, "Locational Value of Distributed Energy Resources," Ernest Orlando Lawrence Berkeley National Laboratory, Berkeley CA, 34545, Feb. 2021.
[4] J. E. Contreras-Ocaña, Y. Chen, U. Siddiqi, and B. Zhang, "Non-Wire Alternatives: An Additional Value Stream for Distributed Energy Resources," *IEEE Trans. Sustain. Energy*, vol. 11, no. 3, pp. 1287–1299, Jul. 2020, doi: 10.1109/TSTE.2019.2922882.
[5] S. Kerscher and P. Arboleya, "The key role of aggregators in the energy transition under the latest European regulatory framework," *Int. J. Electr. Power Energy Syst.*, vol. 134, p. 107361, Jan. 2022, doi: 10.1016/j.ijepes.2021.107361.
[6] S. P. Burger, J. D. Jenkins, C. Batlle, and I. J. Perez-Arriaga, "Restructuring Revisited Part 1: Competition in Electricity Distribution Systems," *Energy J.*, vol. 40, no. 3, Jul. 2019, doi: 10.5547/01956574.40.3.sbur.

[7] E. Haesen, A. D. Alarcon-Rodriguez, J. Driesen, R. Belmans, and G. Ault, "Opportunities for active DER management in deferral of distribution system reinforcements," in *2009 IEEE/PES Power Systems Conference and Exposition*, 2009, pp. 1–8. doi: 10.1109/PSCE.2009.4839997.

[8] K. Bruninx, H. Pandzic, H. Le Cadre, and E. Delarue, "On the Interaction Between Aggregators, Electricity Markets and Residential Demand Response Providers," *IEEE Trans. Power Syst.*, vol. 35, no. 2, pp. 840–853, Mar. 2020, doi: 10.1109/TPWRS.2019.2943670.

[9] H. Chao, "Incentives for efficient pricing mechanism in markets with non-convexities," *J. Regul. Econ.*, vol. 56, no. 1, pp. 33–58, Aug. 2019, doi: 10.1007/s11149-019-09385-w.

[10] M. A. Crew, C. S. Fernando, and P. R. Kleindorfer, "The theory of peak-load pricing: A survey," *J. Regul. Econ.*, vol. 8, no. 3, pp. 215–248, Nov. 1995, doi: 10.1007/BF01070807.

[11] H. Chao, "Peak Load Pricing and Capacity Planning with Demand and Supply Uncertainty," *Bell J. Econ.*, vol. 14, no. 1, p. 179, 1983, doi: 10.2307/3003545.

[12] B. S. Lecinq and M. D. Ilic, "Peak-load pricing for electric power transmission," in *Proceedings of the Thirtieth Hawaii International Conference on System Sciences*, Jan. 1997, vol. 5, pp. 624–633 vol.5. doi: 10.1109/HICSS.1997.663225.

[13] M. Prica and M. D. Ilic, "Peak-load pricing based planning for distribution networks under change," in *2006 IEEE Power Engineering Society General Meeting*, Montreal, Que., Canada, 2006, p. 6 pp. doi: 10.1109/PES.2006.1709296.

[14] S. Junlakarn and M. Ilić, "Provision of Differentiated Reliability Services Under a Market-Based Investment Decision Making," *IEEE Trans. Smart Grid*, vol. 11, no. 5, pp. 3970–3981, Sep. 2020, doi: 10.1109/TSG.2020.2986651.